\documentclass[conference]{IEEEtran}
\IEEEoverridecommandlockouts
\usepackage{cite}
\usepackage{amsmath,amssymb,amsfonts}
\usepackage{algorithmic}
\usepackage{graphicx}
\usepackage{textcomp}
\usepackage{xcolor}
\usepackage{comment}
\usepackage{subfig}
\usepackage{xurl}
\usepackage{tabularx}
\usepackage{babel}
\usepackage{multirow}
\usepackage{color}
\def\BibTeX{{\rm B\kern-.05em{\sc i\kern-.025em b}\kern-.08em
    T\kern-.1667em\lower.7ex\hbox{E}\kern-.125emX}}

\begin{document}

\title{
Optimal Energy Rationing for Prepaid Electricity Customers
\\
\thanks{Maitreyee Marathe was supported by the George Bunn Wisconsin Distinguished Graduate Fellowship provided by the University of Wisconsin-Madison. Line Roald acknowledges support from the U.S. National Science Foundation under Award Number ECCS-2045860.
}
}

\author{\IEEEauthorblockN{Maitreyee Marathe and Line A. Roald}
\IEEEauthorblockA{\textit{Dept. of Electrical and Computer Engineering,}
\textit{University of Wisconsin-Madison,}
Madison, USA \\
\{mmarathe, roald\}@wisc.edu
}
}

\maketitle

\begin{abstract}
For a large (and recently increasing) number of households, 
affordability is a major hurdle in accessing sufficient electricity and avoiding service disconnections. 
For such households, in-home energy rationing, i.e. the need to actively prioritize how to use a limited amount of electricity, is an everyday reality.
In this paper, we consider a particularly vulnerable group of customers, namely prepaid electricity customers, who are required to pay for their electricity a-priori.
With this group of customers in mind, we propose an optimization-based energy management framework to effectively use a limited budget and avoid the disruptions and fees associated with disconnections. 
The framework considers forecasts of future use and knowledge of appliance power ratings to help customers prioritize and limit use of low-priority loads, with the goal of extending access to their critical loads. Importantly, the proposed management system has minimal requirements in terms of in-home hardware and remote communication, lending itself well to adoption across different regions and utility programs. 
Our case study demonstrates that by considering both current and future electricity consumption and more effectively managing access to low-priority loads, the proposed framework increases the value provided to customers and avoids disconnections.  
\end{abstract}

\section{Introduction}

Reliability in the electric power systems literature typically focuses on ensuring that the infrastructure is able to supply electricity to customers on demand. However, for many low-and-moderate income (LMI) households, a main reason for ``power outages'' is inability to pay electric bills, which may prompt service disconnections.
To avoid disconnections, LMI customers are known to ration energy use and compromise on critical needs to pay for energy - referred to as the ``heat or eat" problem \cite{bhattacharya2003heat}. A survey showed that 20\% households in the United States reduced or forwent food or medicine to pay energy costs in 2020 \cite{recs}. 
A particularly vulnerable group is customers in prepaid programs, who have to purchase credits for the energy prior to use, similar to pay-as-you-go phones. These programs are often targeted towards LMI customers, who may be enrolled either voluntarily or forcefully \cite{2022_citizensadvice}. While prepaid programs have some advantages, such as the flexibility to make multiple small payments during the month and avoiding upfront credit checks or deposits, they also present some significant disadvantages. For example, prepaid customers may pay a higher price for electricity \cite{2022_citizensadvice}. Whereas utilities are required to inform postpaid customers before a disconnection, prepaid customers can be automatically and immediately disconnected if their credit runs out. These unanticipated disconnections can be dangerous during extreme heat or cold events and for medically fragile customers. Furthermore, each disconnection-reconnection event can have a fixed charge as high as \$75 \cite{2014_liheap}. 
Despite those disadvantages, it is estimated that there are between 1 to 2.5 million prepaid electricity accounts in the United States \cite{texas_hostage_nodate}, \cite{noauthor_smart_2017} and several million in the United Kingdom \cite{2022_financialtimes}. 

Considering that LMI households form a large part of the customer base of prepaid programs, many prepaid customers may not be able to refill their prepaid accounts with an amount commensurate with their desired use. This presents a need for an effective energy rationing and management framework for home appliances for this customer group, a topic that has received little attention to date. The underlying assumptions in energy rationing are significantly different from demand response, which typically assumes that changes in electricity use should be nearly invisible and non-disruptive to the customer. It is also different from typical literature on 
home energy management systems (HEMS) \cite{beaudin2015home}, which often rely on specific in-home communication, computing, and switching hardware that may not be affordable for LMI families.
%
Aspects of prepaid electricity service have been addressed in terms of meter technology \cite{narmada2017design,gaur2014hdl}, power theft \cite{parvin2015framework}, cyber security \cite{anderson2010controls} and data management \cite{parvin2012standard}. However, only few studies explicitly account for a user-defined budget \cite{ismail2021mixed} or address the home energy management problem specifically for prepaid customers \cite{souza2020automatic}. The former study generates a time-based schedule for load actuation and switches each load ON/OFF, which may be experienced as intrusive, while the latter needs detailed information about operation time per load. 

In contrast,
we present an energy rationing framework for prepaid electricity customers based on a simple threshold-based load control scheme adopted from a DC microgrid setting \cite{manur2020distributed}. Rather than directly actuating loads, 
this framework \emph{enables} the use of a load by comparing the money available in the prepaid account to a monetary threshold assigned to the load. We refer to the prepaid account as the \emph{wallet} and the amount of money in the account as the \emph{wallet balance} or just \emph{balance}. If the balance is higher than the threshold, the load is enabled, i.e. it can be turned on if desired. Otherwise, it should remain off. The benefit of the proposed method is that the available budget may be more evenly spread throughout the month, and high priority loads (with lower thresholds) can be used longer. Furthermore, all loads are turned off before a disconnection occurs, thus avoiding potentially expensive disconnection-reconnection events. 

While such a threshold-based method is very easy to implement, it requires careful definition of the enabling thresholds. Thus, we formulate an optimization problem that identifies optimal threshold values while accounting for the available budget and forecasts of future electricity use. We assume that this problem is solved at regular, yet relatively infrequent intervals (e.g. daily), such that the approach lends itself well to implementation with existing 
in-home hardware and minimal remote communication.

In summary, this paper has two main contributions. First, we design an energy management framework that incorporates user-defined load priorities and helps users optimally ration their allocated energy budget across uses, while requiring minimal local computing and remote communication. This framework builds upon the control framework for prosumers in DC microgrids presented in \cite{manur2020distributed}, but extends it to (i) consider the prepaid wallet balance as a metric of available energy (ii) optimize the activation thresholds for individual loads to achieve better user satisfaction. Second, we implement our proposed method in code and demonstrate its benefit for managing household loads in a case study based on real-life energy use data obtained through the Pecan Street Dataset \cite{noauthor_dataport_nodate}.

The remaining paper is organized as follows.
Section \ref{section:model} presents the energy management framework, the optimization formulation, and comments on practical implementation. Section \ref{section:case-study} presents a case study and numerical results, while Section \ref{section:conclusion} summarizes and concludes the paper.
\section{Model}
\label{section:model}

In this section, we provide details on the model setup and the formulation of the optimal energy rationing problem. 

\subsection{Threshold-based energy management}

The proposed energy rationing framework is an extension of the Self-Organizing Local Electrical Energy Network (SOLEEN), a control framework developed for DC prosumer microgrids without central control or communication \cite{manur2020distributed}. In the original SOLEEN design, each entity or household in the microgrid is assumed to have battery energy storage, with the battery state of charge representing available energy. Load control is done 
by assigning each load a threshold in terms of the state of charge of the battery. If the state of charge falls below the threshold, the corresponding load is switched off. When it is above the threshold, the load is enabled, i.e. the user can decide to switch it on or off. 

Here, we adapt the SOLEEN methodology for a prepaid customer, who may not have a battery. 
Instead, we treat the prepaid wallet balance as a measure of the available energy and express load thresholds in terms of \$ (instead of state of charge) for each load. If the \$ balance falls below this threshold, the load is disabled. 
Due to this analogy with battery energy storage, we refer to the refilling of the wallet (in \$) as a ``recharge". A major difference between a battery and a prepaid wallet is that the battery has a fixed capacity and the state of charge cannot go beyond that, whereas a prepaid wallet does not have an upper limit. Furthermore, electricity demand typically follows a diurnal pattern and the battery capacity (if recharged daily with e.g. solar PV panels) acts as a daily energy budget. In the case of a prepaid wallet, the user would not typically recharge daily. To incorporate a daily limit, we define a \emph{virtual wallet} which is different from the actual prepaid wallet, which we will refer to as the \textit{real wallet}. To match the daily pattern of electricity usage, we recharge the virtual wallet at the beginning of each day with a daily budget and compare load \$ thresholds to the virtual wallet balance. 
This setup brings in a daily budget for energy expenditure and assumes that a user would rather use a small amount of energy every day till the next recharge than use a large amount for the first few days till the wallet runs out and none for the remaining days.

\begin{figure}
\centerline{\includegraphics[width = 0.35\textwidth]{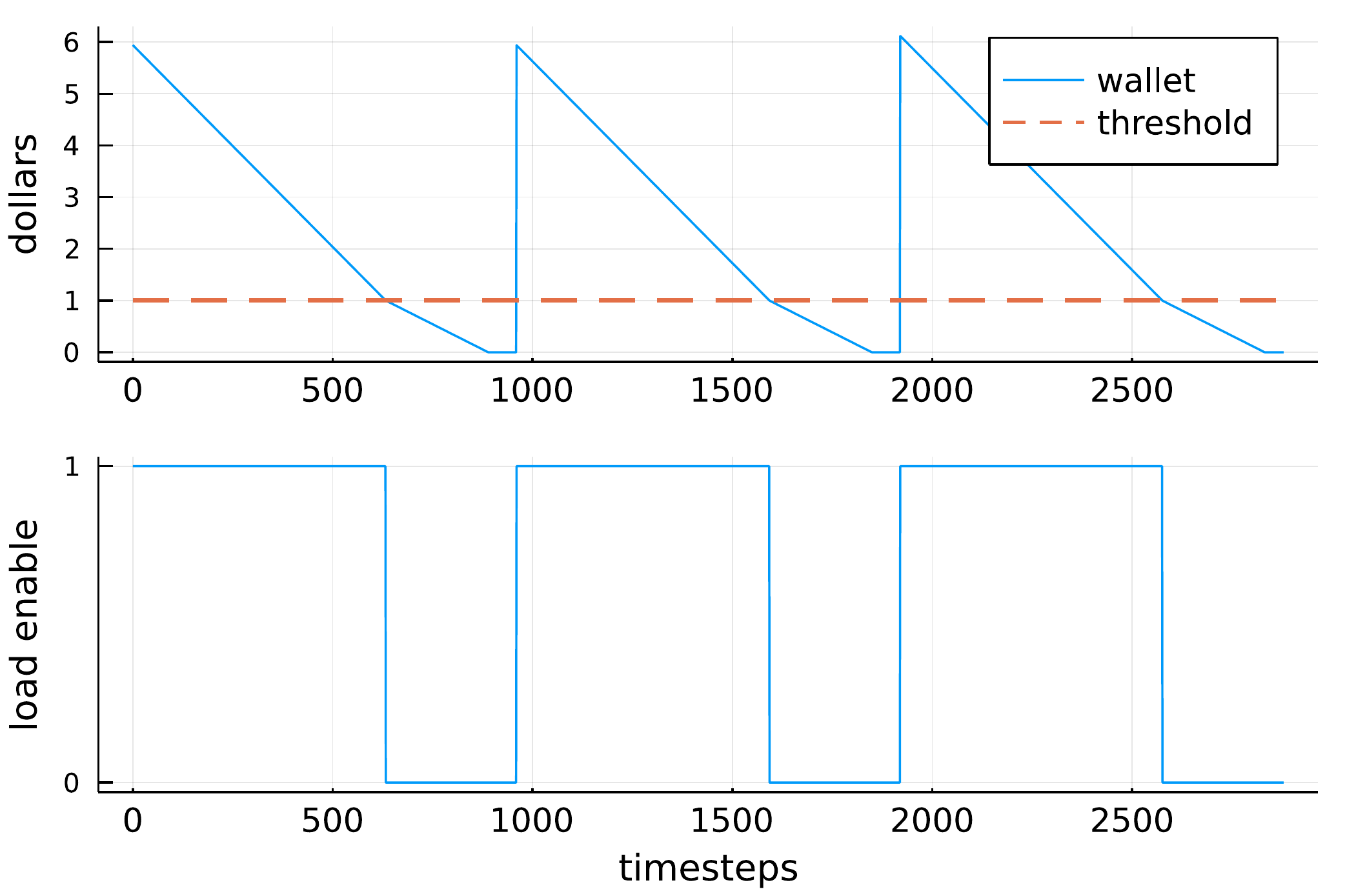}}
\caption{\small Top: Wallet balance (blue) and load threshold (red) in \$. Bottom: Enable signal for the load. A value of 1 indicates that the load can be turned on, while a value of 0 indicates that the load should remain off. 
\vspace{-5mm}}
\label{fig:illustrative-example}
\end{figure}

\subsubsection{Illustrative example} Figure \ref{fig:illustrative-example} shows an example of the proposed framework. The upper plot shows the balance in the virtual wallet along with the threshold for one of the loads, while the lower plot shows whether or not this load is enabled. 
In this example, the virtual wallet is refilled three times (at timestep 0, 960, 1920).  As the wallet balance drops below the threshold, the load is disabled. After this, the considered load does not consume any power and the balance of the virtual wallet reduces less quickly (other loads with a higher priority continue to consume power, such that the wallet balance continues to decrease until it reaches zero and all loads are disconnected).
Once the virtual wallet is recharged, the balance is above the threshold and the load is again enabled. 

\subsubsection{Determining control variables}
With the threshold-based control framework, there are two main control variables that need to be determined, namely (1) how the virtual wallet is recharged as a function of the real wallet balance and (2) how to choose thresholds for effective energy rationing. 
To determine the daily virtual wallet recharge, we simply divide the most recent recharge amount for the real wallet by the number of days until the next recharge\footnote{We assume perfect information about when and what amount will be added to the real wallet. The effect of imperfect real recharge information may be addressed by more sophisticated daily virtual wallet recharge schemes, such as allocating a variable daily budget based on the fraction of the average real recharge per month that has been spent. This is a part of future work.}. This uniformly distributes the latest recharge amount across all days till the next recharge. 
The process for determining the load enable thresholds is more sophisticated. These thresholds are obtained by solving an optimization problem as discussed below.

\vspace{-1.3mm}
\subsection{Optimization formulation}
\label{section:optimization-formulation}
The optimization model determines thresholds for each load $k$ at time $t$ denoted by $\mathbf{x_{k,t}}$. The thresholds for each load are kept constant throughout each day, so the number of unique thresholds is equal to the number of loads times the number of days in the optimization horizon. The input parameters include user-defined priority order for loads, demand forecast, and recharge schedule forecast. We implement the model as a rolling horizon problem with a forecast horizon of seven days and a time step of 15 minutes. We solve the problem once daily, then use the optimized thresholds for the first day to simulate the use of loads to calculate the resulting real and virtual wallet balances. These balances, along with updated forecasts, are used as an input when we re-solve the optimization problem the next day. 
The resulting optimization problem is a mixed-integer linear programming problem.
The nomenclature is given in Table \ref{tab:nomenclature}. 

\begin{table}[]
    \centering    
    \caption{Nomenclature}
    \begin{tabular}{ll}
        \multicolumn{2}{l}{\textbf{Parameters}}\\
        $\mathcal{K}$ & Set of all loads\\
$\mathcal{T}$ & Set of all time steps in the horizon\\
$\Delta T \in \mathbb{R}_{> 0}$ & Duration of time step in  hours\\
$m \in \mathbb{R}_{<0}$ & Large magnitude constant \\
$M \in \mathbb{R}_{>0}$ & Large magnitude constant\\
$\epsilon \in \mathbb{R}_{>0}$ & Small magnitude constant \\
$\alpha \in \mathbb{R}_{>0}$ & Electricity rate in \$/Wh\\
$\gamma_k \in \mathbb{R}_{>0}$ & Priority factor for load $k$ \\
$P_{k,t} \in \mathbb{R}_{\geq 0}$ & Demand in W for load $k$ at time $t$, \\
$d_{k,t} \in \{0,1\}$ & Indicator parameter of demand \\
& $d_{k,t} = 1$ if $P_{k,t} > 0$, and $0$ otherwise\\
$Z_t \in \mathbb{R}_{\geq 0}$ & Real wallet recharge at time $t$ in \$\\
$X_t \in \mathbb{R}_{\geq 0}$ & Virtual wallet recharge at time $t$ in \$ \\[+3pt]
        \multicolumn{2}{l}{\textbf{Variables}}\\
        $\mathbf{z_t} \in \mathbb{R}$ & Real wallet balance at time $t$\\
$\mathbf{u^z_{k,t}} \in \{0,1\}$ & Real enable signal for load $k$ at time $t$\\
$\mathbf{x_t} \in \mathbb{R}$ & Virtual wallet balance at time $t$\\
$\mathbf{x_{k,t}} \in \mathbb{R}$ & Threshold for load $k$ at time $t$\\
$\mathbf{u^x_{k,t}} \in \{0,1\}$ & Virtual enable signal for load $k$ at time $t$ \\ 
$\mathbf{a_{k,t}} \in \{0,1\}$ & Actuation state of load $k$ at time $t$
    \end{tabular}
    \label{tab:nomenclature}
\vspace{-4mm}
\end{table}


\subsubsection{Real wallet constraints}
The amount of money in the real wallet gets updated according to the energy consumption in the previous time step and the recharge amount scheduled for the current time step as given by
\begin{equation}
    \mathbf{z_{t}} = \mathbf{z_{t-1}} + Z_t - \alpha\Delta T \sum_{k\in\mathcal{K}}(P_{k,t-1}\mathbf{a_{k,t-1}}) \quad \forall t \in \mathcal{T}
\label{eq:real-recharge}
\end{equation}
\vspace{-1mm}
Here, $Z_t$ is equal to the recharge scheduled for the day (if any) if $t$ is the first time step of the day and zero otherwise. 
The real enable signal $\mathbf{u^z_{k,t}}$ expresses whether there is money in the real wallet. If the real wallet balance $\mathbf{z_t}$ is positive, $\mathbf{u^z_{k,t}}=1$, and otherwise is zero. This is enforced by
\begin{align}
    m\mathbf{u^z_{k,t}}&\leq -\mathbf{z_t} \quad &&\forall k \in \mathcal{K}, \forall t \in \mathcal{T}
    \label{eq:real-enable}\\
    (M+\epsilon)(1-\mathbf{u^z_{k,t}})&\geq \epsilon - \mathbf{z_t} \quad &&\forall k \in \mathcal{K}, \forall t \in \mathcal{T}
    \label{eq:real-disable}
\end{align}

\subsubsection{Virtual wallet constraints}
The virtual wallet balance $\mathbf{x_t}$ is updated according to the energy consumption since the previous time step and the daily virtual recharge $X_t$ computed from the real recharge, i.e.
\begin{equation}
    \mathbf{x_t} = \mathbf{x_{t-1}} + X_t - \alpha\Delta T\sum_{k}(P_{k,t-1}\mathbf{a_{k,t-1}}) \quad \forall t \in \mathcal{T}
    \label{eq:virtual-recharge}
\end{equation}
where $X_t$ is equal to the recharge scheduled for the day if $t$ is the first time step of the day and zero otherwise. 
The virtual enable signal $\mathbf{u^x_{k,t}} \in \{0,1\}$ 
should be 1 if the virtual wallet balance $\mathbf{x_t}$  is greater than or equal to the load threshold $\mathbf{x_{k,t}}$, or otherwise zero, as expressed by
\begin{align}
    \mathbf{x_t} - \mathbf{x_{k,t}} + \epsilon &\leq (M+\epsilon)\mathbf{u^x_{k,t}} \quad &&\forall k \in \mathcal{K}, \forall t \in \mathcal{T}
    \label{eq:virtual-enable}\\
    \mathbf{x_t} - \mathbf{x_{k,t}} &\geq m(1-\mathbf{u^x_{k,t}}) \quad &&\forall k \in \mathcal{K}, \forall t \in \mathcal{T}
    \label{eq:virtual-disable}
\end{align}
Note that $\mathbf{x_{k,t}}$ is also constrained to be the same for all time steps $t$ during a single day.


\subsubsection{Actuation constraints}
The actuation constraints describe whether a load is on and consuming power. 
The actuation state $\mathbf{a_{k,t}}$ of load $k$ at time $t$ is zero if there is no demand $d_{k,t}$, the virtual wallet balance $\mathbf{x_t}$ is less than its threshold $\mathbf{x_{k,t}}$, if the real wallet balance $\mathbf{z_t}$ is less than or equal to zero, or if the real wallet balance in the next time step $\mathbf{z_{t+1}}$ would be less than or equal to zero if the load is kept on (the latter condition ensures that the real wallet balance does not go negative by the next time step). In this case, $\mathbf{a_{k,t}}=0$ as expressed by the following constraints
\begin{equation}
    \mathbf{a_{k,t}}\!\leq\! d_{k,t}\mathbf{u^x_{k,t}}, ~~ \mathbf{a_{k,t}}\!\leq\!\mathbf{u^z_{k,t}}, ~~ \mathbf{a_{k,t}} \!\leq\!\mathbf{u^z_{k,t+1}} ~~\forall k\!\in\!\mathcal{K}, \forall t\!\in\!\mathcal{T} \nonumber
\end{equation}
If none of the above conditions are satisfied, the actuation state has to be equal to 1, i.e. $\mathbf{a_{k,t}}=1$, as described by 
%
\begin{equation}
    d_{k,t}\mathbf{u^x_{k,t}} + \mathbf{u^z_{k,t}} + \mathbf{u^z_{k,t+1}} \leq 2 + \mathbf{a_{k,t}} \quad \forall k \in \mathcal{K}, \forall t \in \mathcal{T}
    \label{eq:actuation-on}
\end{equation}

\subsubsection{Objective function}
Our goal is to maximize the value provided to the customer from using a limited (less than desired) amount of electricity. 
To express this, we make a few key assumptions.
First, some loads bring more value, i.e. 
have a higher priority, as compared to others and the value of a load does
not necessarily scale with its power rating. For instance, a light bulb rated at 5W can provide more value than a TV rated at 100W. Second, we assume that value is associated with whether or not a load is available when desired. Therefore, we design our objective function to account for (1) the priority assigned to each load by the user and (2) the percentage of time the load was available when desired. 

We introduce the \emph{load priority factor} $\gamma_k$ to express the relative value of satisfying the demand of load $k$ compared to other loads. It is calculated as $\gamma_k = \frac{1}{\eta_k}\frac{1}{\sum\limits_k \frac{1}{\eta_k}}$, where $\eta_k$ is a number that represents the position of the load in the priority order (lower numbers imply higher priority). For example, if the user ranks loads k = 1, 2, 3 in the order 2, 3, 1 then $\eta_1 = 2$, $\eta_2 = 3$, $\eta_3 = 1$. Note that this is not the only way to determine load priority factor, and this was chosen since it ensures that $\sum_k \gamma_k = 1$. 

Next, we define the \emph{Service Factor} ($SF_k$) for each load $k$ as the ratio of the amount of time a load was available to the amount of time it was demanded.
\begin{equation}
    SF_k = {\sum\limits_t \mathbf{a_{k,t}}}~{\huge /}~{\sum\limits_t d_{k,t}}
    \label{eq:service-factor-definition}
\end{equation}
The service factor expresses the percentage of time the load was available when desired and is independent of the power rating of the load. For example, a light rated at 5W demanded for 6 hours and served for 3 hours will have the same service factor as a TV rated at 100W demanded for 2 hours and served for 1 hour. Both will have $SF_k = 0.5$. 

The \emph{Priority Service Factor} (PSF) is the weighted average of the service factors, where the weight is the load priority factor. Our objective is to maximize the PSF, 
\begin{equation}
    \max~~ \sum_{k\in\mathcal{K}}\gamma_{k}SF_k
    \label{eq:psf}
\end{equation}

\subsection{Benchmark Cases}
To compare the proposed method, we also implement a simulation of two benchmark cases:
\subsubsection{Baseline} The baseline case does not have any energy manager and assumes that all loads are enabled as long as the real wallet balance is positive (i.e., it is a greedy approach). From a mathematical perspective, this formulation only considers the real wallet update \eqref{eq:real-recharge} and sets $\mathbf{a_{k,t}}=1$ if $d_{k,t}=\mathbf{u_{k,t}^z}=1$. 
\subsubsection{Fixed Thresholds} The fixed thresholds case uses a similar model as the one described in Section \ref{section:optimization-formulation}, except that the load thresholds $\mathbf{x_{k,t}}$ are fixed (i.e., not optimized). While there could be many methods for computing these fixed thresholds, we choose to calculate them once at the beginning of the month as 
$\mathbf{x_{k,t}} = \frac{\eta_k}{N} \beta X \quad \forall t$.
Here, $\eta_k$ is the position of Load $k$ in the priority order, $N$ is the total number of loads, $X$ is the total recharge amount for the month (in \$), and $\beta$ is a positive constant less than 1. For this study, $\beta = 0.05$ was used. With this definition, lower priority loads (i.e. those with a higher $\eta_k$) are assigned higher thresholds and will get disabled before higher priority loads as the wallet balance drops. 

\subsection{Implementation through a prepaid program}
A benefit of the proposed framework is that it requires minimal additional hardware, computation and remote communication, and thus lends itself well to implementation using the existing infrastructure without significant technological overheads. 

\noindent\emph{A) Computational requirements:} Prepaid programs generally install in-home displays in the customer's house that show their wallet balance \cite{noauthor_srp_nodate}. 
The in-home display is connected to the meter and computes and displays the real wallet balance in real-time. It can be programmed to keep track of the virtual wallet as well, and since this is a simple algebraic operation, additional computation hardware may not be necessary. 

\noindent\emph{B) Communication requirements:} Households that struggle to pay energy bills may also fall behind on internet and phone bills, thus limiting options for communication. The optimized thresholds can be computed on a remote server and communicated to the in-home display at the beginning of each day through the same communication channel that the utility uses to connect to the in-home display (e.g. the Advanced Metering Infrastructure (AMI) communication system). Since thresholds need to be communicated only once a day, communication delays can be tolerated. 

\noindent\emph{C) Hardware requirements:} Direct actuation of loads in a user's home requires additional hardware such as smart switches. If users cannot afford this additional hardware or are not comfortable with an application directly controlling certain household appliances such as medical equipment,
either the in-home display or a phone application can instead provide ``nudges'' to suggest to the user that they switch off the load to allow them to use higher priority loads later on.

\section{Case study}
\label{section:case-study}
We next demonstrate our proposed framework for optimal energy rationing with a case study. We compare the proposed optimal method against the two benchmark cases, and assess the impact of recharge frequency and overall recharge amount on the priority service factor and the number of disconnections.

\begin{figure}
\centerline{\includegraphics[width = \columnwidth]{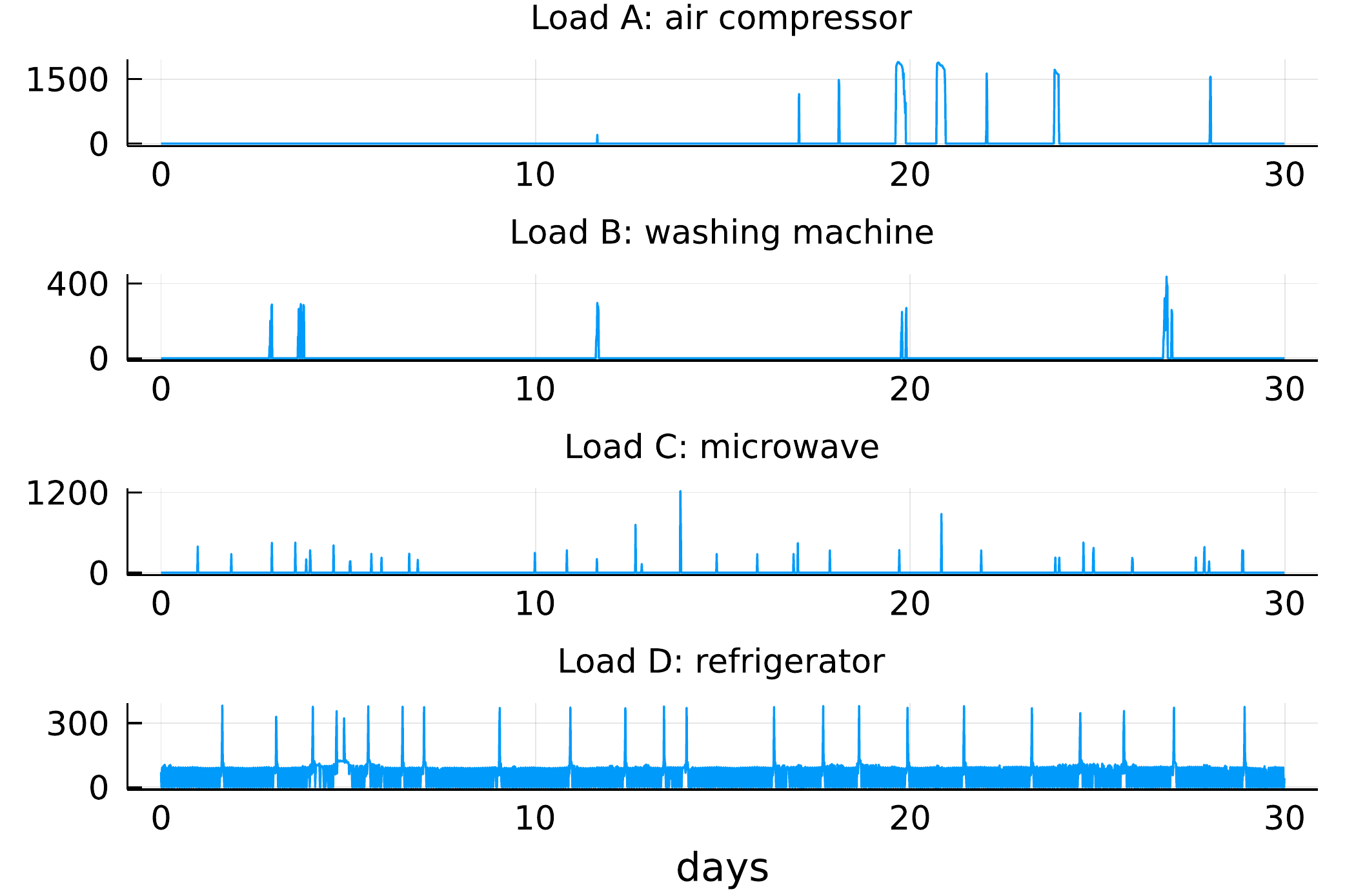}}
\vspace{-2mm}
\caption{Load power demand for 30 days (W).}
\label{fig:load_power_plot}
\end{figure}
\vspace{-3mm}
\begin{table}
\caption{Load data\vspace{-3mm}}
\begin{center}
\begin{tabular}{m{2em} m{7em} m{3em} m{5em} m{3.5em}}
\hline
\multicolumn{2}{l}{\textbf{Appliance}} & \textbf{Energy (kWh)} & \textbf{Max Power (kW)} & \textbf{Load Priority}\\
\hline
A & air compressor & 28 & 1.9 & 2\\ 
B & washing machine  & 2.0 & 0.44 & 4\\
C & microwave & 3.6 & 1.2 & 3\\
D & refrigerator & 41 & 0.38 & 1\\
\hline
\end{tabular}
\vspace{-5mm}
\label{tab:load-data}
\end{center}
\end{table}
\subsection{Setup and implementation}
We use load data for one month (30 days) from one house in the Pecan Street Dataset \cite{noauthor_dataport_nodate}. The four loads used are an air compressor (Load A), a washing machine (Load B), a microwave (Load C), a refrigerator (Load D). Figure \ref{fig:load_power_plot} shows the power demand for each load for the duration of the month and Table \ref{tab:load-data} summarizes the total energy demand, maximum power demand, and priority for each load.

We assume that perfect information about the current and future demand for each load as well as the recharge amount and timing is available to the model. (Considering effects of imperfect information and user non-compliance are a part of future work and may be addressed through more sophisticated schemes such as variable daily virtual wallet recharge.) The presented results assume either direct load actuation through smart switches or total user compliance to the load enable signal nudges given by the model, i.e. we assume that a load is only in use when the corresponding load actuation signal is $\mathbf{a_{k,t}=1}$. We use an electricity rate of $\alpha = \text{\$}0.15/kWh$. Since this is a parameter, time-of-use pricing (i.e. time-varying $\alpha(t)$) may be incorporated without changing the complexity of the problem. The total cost of electricity for using the devices in Table \ref{tab:load-data} with a constant $\alpha = \text{\$}0.15/kWh$ would thus be \$11.19 \footnote{ Additionally, a transaction cost (fee associated with each recharge) and a monthly fixed cost can be included and is considered part of future work. Factors such as whether the fixed cost will be deducted from the wallet per day or upfront at the beginning of the month may influence model performance.}.

The \emph{recharge frequency} is the number of times a user recharges their wallet in a month, and typically varies between 1 and 7 per month \cite{2010_epri}. We express the total recharge as the fraction of the total cost of desired electricity. The \emph{recharge amount} is the total amount of money added to the real wallet over the course of the month. We express this amount as a percentage of the amount needed to cover the desired electricity use listed in Table \ref{tab:load-data}. Thus, recharge amounts $<100$\% imply a need for rationing. 
We assume that the total recharge amount is uniformly distributed across all recharges.

The proposed optimization model and the benchmarking cases are implemented in the Julia programming language \cite{bezanson2017julia} (v1.6) and run on a machine with an Intel CPU @3.2GHz and 16GB memory. In the optimized thresholds case, the simulation calls the optimization model implemented using JuMP \cite{dunning2017jump} and the Gurobi solver \cite{gurobi}. The parameter $\epsilon$ was set to 1e-6 which is equal to the default tolerance of the solver Gurobi to meet constraints. 
The optimization model computes new thresholds daily, using an optimization horizon of 7 days. The thresholds for the current day are then implemented in the simulation. The real and virtual wallet balances at the end of the simulated day are then used as an input for the optimization problem the next day.

\subsection{Comparison with Benchmark Cases}
\begin{figure}
\centerline{\includegraphics[width = 0.67\columnwidth]{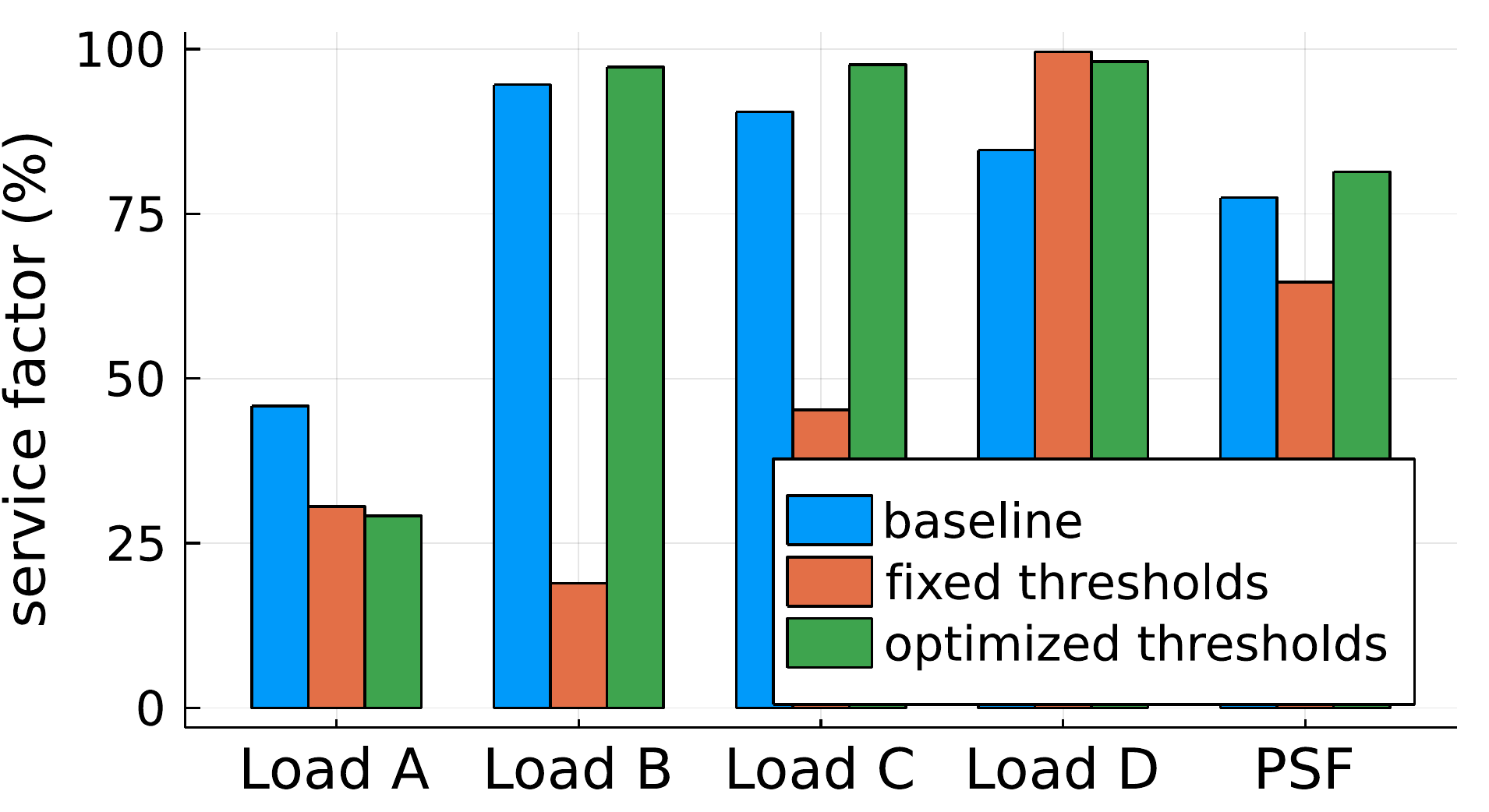}}
\caption{\small Service factor for the 4 loads and overall priority service factor (PSF) at a recharge frequency of 5 per month and recharge amount of 70\%. Load priority: Load$\,$D$\,>\,$Load$\,$A$\,>\,$Load$\,$C$\,>\,$Load$\,$B}
\label{fig:recharge70realrf5load_tsf_psf_barchart}
\end{figure}

As a first investigation, we compare the performance of the optimal energy rationing framework with the baseline and fixed thresholds benchmark cases. To do this, we run each method for the full 30 days and compare the resulting service factors (SF) per load as defined by \eqref{eq:service-factor-definition}, the overall achieved priority service factor (PSF) as defined by \eqref{eq:psf}, as well as the amount of energy consumed per load. For this evaluation, we assume a recharge amount of 70\% and a recharge frequency of 5 payments per month.
Figure \ref{fig:recharge70realrf5load_tsf_psf_barchart} shows the service factor $SF_k$ for each load and the overall PSF and Table \ref{tab:energy-use-per-load} shows the energy use per load. We observe that the different cases lead to very different use of energy across the different loads. The refrigerator (Load D) has the highest priority, and is also the load that runs most continuously. The optimized and fixed threshold cases achieve both a service factor and energy use close to 100\% for this load, while the baseline case (which experiences a prolonged disconnection) only achieves a service factor and energy use of about 85\%. 
The air compressor (Load A) has a power demand that is much higher than other loads and the second highest energy demand. Although this load has the second highest priority, 
the optimized threshold case compromises serving this load to ensure better service to the higher priority Load D and the lower priority but lower energy demand Loads B and C. 
The baseline and fixed thresholds cases serve more energy to the air compressor, but have lower service factors for the microwave (Load C) and the washing machine (Load B).
Because of this, the overall PSF in the optimized thresholds case is greater than that in the other two cases.  
Further, the optimized thresholds case and the baseline case fully use the allocated budget of 70\% of total desired energy, whereas the fixed thresholds case only uses 68\%. This is because the fixed thresholds case always has residual balance equal to at least the lowest threshold i.e. the threshold corresponding to the highest priority load.


\begin{table}
\caption{Energy Usage per Load\vspace{-3mm}}
\begin{center}
\begin{tabular}{ll|l|l|l|l|l|l}
\hline
\multicolumn{2}{l|}{\multirow{3}{*}{\textbf{Appliance}}} & \multicolumn{2}{c|}{\multirow{2}{*}{\textbf{Baseline}}} & \multicolumn{2}{c|}{\textbf{Fixed }} & \multicolumn{2}{c}{\textbf{Optimal}}\\
\multicolumn{2}{c|}{} & \multicolumn{2}{c|}{} & \multicolumn{2}{c|}{\textbf{ Thresholds}} & \multicolumn{2}{c}{\textbf{ Thresholds}}\\[+2pt]
\multicolumn{2}{c|}{}  & \textbf{kWh} & \textbf{\%} & \textbf{kWh} & \textbf{\%} & \textbf{kWh} & \textbf{\%} \\ 
\hline
A & air compressor & 12 & 43 & 7.5 & 27 & 6.7 & 24 \\ 
B & washing machine  & 1.9 & 95 & 0.34 & 17 & 2.0 & 100 \\
C & microwave & 3.2 & 89 & 1.8 & 50 & 3.5 & 97 \\
D & refrigerator & 35 & 85 & 41 & 100 & 40 & 98 \\
\hline
\multicolumn{2}{l|}{\textbf{Total}} & 52.1 & 70 & 50.6 & 68 & 52.2 & 70\\
\hline
\end{tabular}
\vspace{-5mm}
\label{tab:energy-use-per-load}
\end{center}
\end{table}


\subsection{Impact of Recharge Amount and Frequency}
Next, we compare the achieved PSF and the number of disconnections (i.e., how frequently the real wallet balance falls below zero and customers lose electricity supply) across different recharge amounts and frequencies.

\subsubsection{Impact on priority service factor}
\begin{figure}
\centerline{\includegraphics[width = 0.45\textwidth]{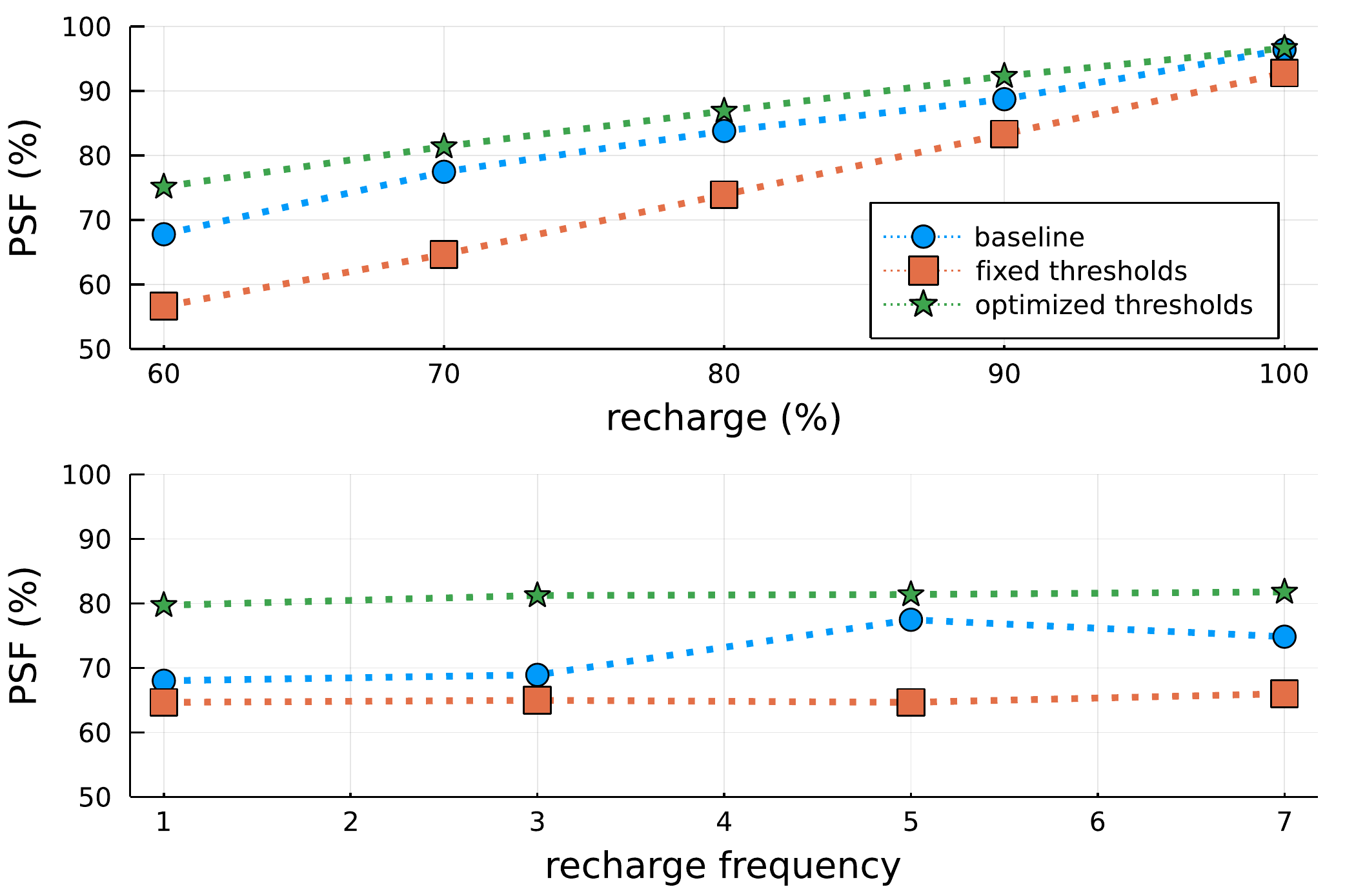}}
\caption{\small Top: PSF against recharge amount at a recharge frequency of 5 per month. 
Bottom: PSF against monthly recharge frequency at a recharge amount of 70\%. 
\vspace{-5mm}}
\label{fig:priority_sf_plot}
\end{figure}

The upper plot in Figure \ref{fig:priority_sf_plot} shows the variation in the PSF as the recharge amount is varied from 60\% to 100\% with a constant recharge frequency of 5 per month. The PSF increases with increasing recharge amount for all three cases, as expected. Across all recharge amounts, the optimized thresholds case has a PSF greater than the baseline while the fixed thresholds case performs worse due to the residual balance that remains unused. The bottom plot in Figure \ref{fig:priority_sf_plot} shows the variation in PSF as the recharge frequency is varied from 1 to 7 per month while keeping the recharge amount constant at 70\%. The optimized and fixed thresholds cases have minimal variation in PSF even if the recharge amount is distributed across increasing number of recharges during the month because the daily virtual wallet recharge remains the same. The optimized thresholds case performs better than the baseline whereas the fixed thresholds case performs worse due to the unused residual energy.

\subsubsection{Impact on disconnections}
We further investigate the impact of recharge frequency and amount on the number of disconnections. We first vary the recharge amount from 60\% to 100\% while keeping the recharge frequency constant at 5 per month, and then vary the recharge frequency from 1 to 7 per month while keeping the recharge amount constant at 70\%.
In all of these cases, both the fixed and optimized threshold cases result in zero disconnections as all loads were disabled before the real wallet balance could fall below zero. This demonstrates that the simple threshold-based management can avoid disconnections and associated reconnection fees.

This is quite an improvement relative to the baseline case. With 5 recharges per month, the baseline case experienced 3 disconnections for recharge amounts of 60\% and 70\% and one disconnection for recharge amounts $\geq 80$
\%.
When we fix the recharge amount at 70\%, we observe that the number of disconnections increase with increasing recharge frequency. With one recharge, we only have one disconnection, with 3 recharges we have 2 disconnections and with 5 and 7 recharges we have 3 disconnections. 
The reason for this behavior is that the recharge amounts are distributed uniformly across the month, whereas the load is not. For low recharge amounts or high recharge frequencies, there is therefore a higher chance that the real wallet balance might drop to zero before the next recharge is made whereas for higher recharge amounts and fewer payments, these intermediate disconnections are avoided. 

\vspace{-3mm}
\section{Conclusion}
\label{section:conclusion}
This work proposes a threshold-based energy rationing framework for prepaid customers. This framework compares a predetermined threshold with the prepaid wallet balance to decide whether a load can be used without impacting other, higher priority loads later. To determine the optimal threshold values, we formulate and solve a rolling horizon optimization problem.
The case study shows that the framework with optimized thresholds outperforms both a baseline without energy management and a method where control thresholds are fixed solely based on load priority information. Specifically, the proposed method serves higher priority loads and reduces disconnections by curtailing lower priority loads.

Some avenues for future work include studying the effects of imperfect forecasts of future demand and recharge frequency and amount. The effect of incorporating fixed and transaction costs can also be studied, and be used to inform overall prepaid program design. Finally, while the framework has been presented in the context of LMI households in the United States, it can be extended for other contexts such as postpaid customers on energy assistance programs and pay-as-you-go solar home system users. 

\bibliographystyle{IEEEtran}
\bibliography{myBibliography}

\end{document}